\documentclass[preprint,aps]{revtex4-1}

\usepackage{graphicx}

\usepackage{soul}
\usepackage{epstopdf}
\usepackage{bm}
\usepackage{upgreek}
\usepackage{amsmath}
\usepackage{lipsum}
\usepackage{color,xcolor}

\begin{document}

\title{Negligible Normal Fluid in Superconducting State of Heavily Overdoped Bi$_2$Sr$_2$CaCu$_2$O$_{8+\delta}$ Detected by Ultra-Low Temperature Angle-Resolved Photoemission Spectroscopy} 

\author{Chaohui Yin$^{1,2,\dag}$, Qinghong Wang$^{1,2,\dag}$, Yuyang Xie$^{1,2}$, Yiwen Chen$^{1,2}$, Junhao Liu$^{1,2}$, Jiangang Yang$^{1,2}$, Junjie Jia$^{1,2}$, Xing Zhang$^{1,2}$, Wenkai Lv$^{1,2}$, Hongtao Yan$^{1,2}$, Hongtao Rong$^{1,2}$, Shenjin Zhang$^{3}$, Zhimin Wang$^{3}$, Nan Zong$^{3}$, Lijuan Liu$^{3}$, Rukang Li$^{3}$, Xiaoyang Wang$^{3}$, Fengfeng Zhang$^{3}$, Feng Yang$^{3}$, Qinjun Peng$^{3}$, Zuyan Xu$^{3}$, Guodong Liu$^{1,2,4}$, Hanqing Mao$^{1,2,4}$, Lin Zhao$^{1,2,4,*}$, Xintong Li$^{1,2,4,*}$ and Xingjiang Zhou$^{1,2,4,*}$}

\affiliation{
	\\$^{1}$National Lab for Superconductivity, Beijing National Laboratory for Condensed Matter Physics, Institute of Physics, Chinese Academy of Sciences, Beijing 100190, China
	\\$^{2}$University of Chinese Academy of Sciences, Beijing 100049, China
	\\$^{3}$Technical Institute of Physics and Chemistry, Chinese Academy of Sciences, Beijing 100190, China
	\\$^{4}$Songshan Lake Materials Laboratory, Dongguan, Guangdong 523808, China
	\\$^{\dag}$These authors contributed equally to this work.
	\\$^{*}$Corresponding author: XJZhou@iphy.ac.cn, xintongli@iphy.ac.cn, lzhao@iphy.ac.cn	
}

\date{\today}

\maketitle

\newpage

{\bf  In high temperature cuprate superconductors, it was found that in the overdoped region the superfluid density decreases with the increase of hole doping. One natural question is whether there exists normal fluid in the superconducting state in the overdoped region. In this paper, we have carried out high-resolution ultra-low temperature laser-based angle-resolved photoemission measurements on a heavily overdoped Bi2212 sample with a $T_{\mathrm{c}}$ of 48\,K. We find that this heavily overdoped Bi2212 remains in the strong coupling regime with $2 \mathit{\Delta}_0 / k_{\mathrm{B}} T_{\mathrm{c}}=5.8$. The single-particle scattering rate is very small along the nodal direction ($\sim$5\,meV) and increases as the momentum moves from the nodal to the antinodal regions. A hard superconducting gap opening is observed near the antinodal region with the spectral weight at the Fermi level fully suppressed to zero. The normal fluid is found to be negligibly small in the superconducting state of this heavily overdoped Bi2212. These results provide key information to understand the high $T_\mathrm{c}$ mechanism in the cuprate superconductors. 
	
	}

~\\
\noindent{\bf\large Introduction}

The high temperature cuprate superconductors exhibit anomalous normal state properties as well as unconventional superconducting properties.\textsuperscript{\cite{keimer2015quantum}} In the underdoped region, with the increase of the doping level, the superconducting transition temperature ($T_\mathrm{c}$) and the superfluid density increase while the superconducting gap decreases, indicating that superconductivity is closely related to superfluid density.\textsuperscript{\cite{uemura1989universal,hashimoto2014energy}} In the overdoped region, the superconducting transition temperature, the superconducting gap and superfluid density all decrease with the increase of the doping level.\textsuperscript{\cite{bovzovic2016dependence}} One prominent issue arises as to whether there is normal fluid in the superconducting state due to the missing charge carriers if fewer and fewer carriers are condensed into the superconducting state in the overdoped region.\textsuperscript{\cite{mahmood2019locating}} In Bi$_2$Sr$_2$CuO$_{6+\delta}$ (Bi2201) superconductors, normal electrons are observed in the superconducting state by scanning tunneling microscopy (STM) both in the real space\textsuperscript{\cite{tromp2023puddle}} and in the reciprocal space\textsuperscript{\cite{ye2023emergent}} which are strongly affected by disorder. Whether it is universal that the normal fluid always exists in the superconducting state in the overdoped samples remains to be investigated.

Angle-resolved photoemission spectroscopy (ARPES) is a powerful tool to study the normal state and superconducting state of the cuprate superconductors.\textsuperscript{\cite{damascelli2003angle,sobota2021angle}} It can also detect the normal electrons in the superconducting state. Here we report high-resolution ultra-low temperature ARPES studies on heavily overdoped Bi$_2$Sr$_2$CaCu$_2$O$_{8+\delta}$ (Bi2212) with a $T_\mathrm{c}$ at 48\,K. By performing ARPES measurements near the antinodal region at ultra-low temperature (1.2\,K), we find that the normal fluid is negligibly small in the superconducting state in this heavily overdoped Bi2212 sample. These results provide key information for understanding the mechanism of superconductivity in cuprate superconductors.

High-quality Bi2212 single crystals were grown using the traveling solvent floating zone method.\textsuperscript{\cite{liang2002growth,wen2008large}} The crystals were annealed at 500\,${ }^{\circ} \mathrm{C}$ under high oxygen pressure over 300\,bar for seven days.\textsuperscript{\cite{zhang2016reproducible}} Magnetic susceptibility measurement (Fig. 1(a)) indicates that the prepared sample has a $T_\mathrm{c}$ at 48.5\,K which is among the most heavily overdoped samples that can be achieved in bulk Bi2212 (Fig. 1(b)). For convenience, the sample will be referred to as Bi2212OD48K hereafter. High-resolution ARPES measurements were conducted using our laboratory-based ARPES system. It is equipped with a vacuum-ultraviolet laser operating at a photon energy of $h\nu$ =6.994\,eV and a R8000 hemispherical electron energy analyzer (Scienta Omicron). Employing $^{3}\mathrm{He}$ pumping technology, the system can cool samples to the lowest temperature of $\sim$0.8\,K. The energy resolution was set to 0.5\,meV and the angular resolution is 0.3° which corresponds to a momentum resolution of 0.004\,\AA$\mathrm{^{-1}}$ at the photon energy of 6.994\,eV. All samples were cleaved $in$ $situ$ at low temperatures and measured in ultrahigh vacuum with a base pressure better than $\mathrm{1 \times 10^{-10}}$ mbar. The Fermi level was carefully referenced by measuring polycrystalline gold which was well connected with the sample. The work function of the heavily overdoped Bi2212 was measured to be 4.35\,eV. In the STM measurements, the sample was cleaved at room temperature in an ultrahigh vacuum and the data were obtained at 4.7\,K. The tip used in the measurements was an etched tungsten tip which has been calibrated on the Au(111) surface. The differential conductance (d$I$/d$V$) spectra were obtained by using the standard lock-in technique with a modulation of 3\,mV and 488.137\,Hz.

~\\
\noindent{\bf\large{Results and discussion}}

 Figure 1 presents detailed Fermi surface measurements of the Bi2212OD48K sample at 1.2\,K. Figure 1(c) shows the Fermi surface mapping while Figs. 1(d) and 1(e) show the constant energy contours at binding energies of 5\,meV and 10\,meV, respectively. With increasing binding energy, the spectral weight spreads from the nodal region to the antinodal region due to an anisotropic superconducting gap opening. In the constant energy contour with a binding energy of 10\,meV (Fig. 1(e)), in addition to the strong antibonding Fermi surface, the bonding Fermi surface becomes visible near the antinodal region. But it is rather weak due to the photoemission matrix element effect. Combined with the measured band structures (Fig. 2), these results give a Fermi surface topology that is plotted in Fig. 1(f). It consists of the bonding Fermi surface and the antibonding Fermi surface. Analyzing the area of the measured Fermi surface, the doping level of the antibonding Fermi surface is  $\sim$0.36\,hole/Cu while it is $\sim$0.20\,hole/Cu for the bonding Fermi surface. These yield an overall doping level of $\sim$0.28\,hole/Cu, which is slightly higher than 0.24\,hole/Cu estimated from empirical formula (Fig. 1(b)).\textsuperscript{\cite{presland1991general}} We note that, even though the antibonding Fermi surface has a doping level of $\sim$0.36\,hole/Cu, it remains hole-like centered around ($\pi$,$\pi$). This indicates that even up to this heavily overdoped Bi2212 sample, the antibonding Fermi surface has not undergone a Lifshitz transition from a hole-like to an electron-like topology, as reported before. 
 \textsuperscript{\cite{kaminski2006change}}
 
 Figure 2 shows the band structures of the Bi2212OD48K sample measured at 1.2\,K. The antibonding band is strong and clear in all momentum cuts. The bonding band appears clearly near the antinodal region although its intensity is relatively weak. They split more and more when the momentum cuts approach the antinodal region. The spectral weight near the Fermi level gets increasingly suppressed as the momentum cuts move from the nodal to the antinodal regions. Signatures of the Bogoliubov bands for both the bonding and antibonding bands, caused by the superconducting gap opening, are clearly seen.

 Figure 3 shows the representative photoemission spectra of the Bi2212OD48K sample measured at 1.2\,K along the antibonding Fermi surface. Fig. 3(a) shows the original photoemission spectra (energy dispersion curves, EDCs) at five Fermi momenta and corresponding background EDCs are also plotted. The observed EDCs are very sharp with a width (full width at half maximum, FWHM) of $\sim$5\,meV for the nodal EDC (leftmost panel in Fig. 3(a)) and $\sim$12\,meV for the EDC near the antinodal region (rightmost panel in Fig. 3(a)). The nodal EDC in our heavily overdoped Bi2212 sample is among the sharpest spectra so far observed in Bi-based cuprates. This indicates the scattering rate along the nodal direction is extremely small which corresponds to a long quasiparticle lifetime, suggesting that the material is subject to minimal scattering from disorder or other defects. 
 
 To quantitatively analyze the superconducting gap, scattering rate and the fraction of the normal fluid in the superconducting state, the EDCs along the Fermi surface in Fig. 3(a) are symmetrized as shown in Fig. 3(b). We also subtracted the corresponding background (BG) from the EDCs along the Fermi surface in Fig. 3(a) and the symmetrized background-subtracted EDCs are shown in Fig. 3(c). By subtracting the background, the EDC intensity near the antinodal region approaches zero at the Fermi level, as shown in the insets of the right three panels in Fig. 3(c). The hard gap opening is observed where the spectral intensity is nearly zero within an energy range around the Fermi level. The symmetrized EDCs are fitted by a phenomenological gap formula,\textsuperscript{\cite{norman1998}} taking the self-energy $\mathit{\Sigma}(k, \omega)=-i \mathit{\Gamma}_1+\mathit{\Delta}^2 /[\omega+\epsilon(k)+i\mathit{\Gamma}_0]$ where  $\mathit{\Gamma}_1$ is a single-particle scattering rate, $\mathit{\Gamma}_0$ is the inverse pair lifetime, $\mathit{\Delta}$ is the superconducting gap and $\epsilon(k)$ represents the band dispersion which is zero at the Fermi momentum. This fitting procedure is applied to all the symmetrized background-subtracted EDCs along the antibonding Fermi surface and the fitting results are presented in Fig. 4.
 
 Figure 4(a) shows the superconducting gap measured along the antibonding Fermi surface. The superconducting gap can be well fitted by a standard d-wave gap form $\mathit{\Delta} = \mathit{\Delta}_0 |\cos (k_x a)-\cos (k_y a)| / 2$ where $\mathit{\Delta}_0$ equals 12.0\,meV. 
 This is consistent with our STM measurement on the same sample (bottom-right inset in Fig. 4(a)) where the coherence peak lies at 12.5\,meV. 
 This results in a ratio of the gap to the critical temperature $2 \mathit{\Delta}_0 / k_{\mathrm{B}} T_{\mathrm{c}}=5.8$. It is significantly larger than the value of 4.3 for the weak coupling d-wave BCS superconductor.\textsuperscript{\cite{he2018rapid}} This indicates that the heavily overdoped Bi2212 up to $T_\mathrm{c}\sim$48\,K remains in the strong coupling regime.
 
 Figure 4(b) presents the momentum-dependent single-particle scattering rate ($\mathit{\Gamma}_1$) and inverse pair lifetime ($\mathit{\Gamma}_0$) along the antibonding Fermi surface. $\mathit{\Gamma}_0$ is quite small (less than 1.5\,meV) across the entire Fermi surface.
 The single-particle scattering rate is minimal ($\sim$5\,meV) along the nodal direction, and slightly increases as the momentum moves to the antinodal region and shoots up near the antinodal region. The antinodal electrons appear to experience some additional scattering channels. Previous ARPES measurements found that, in the overdoped (La$_{2-x}$Sr$_x$)Cu$_2$O$_{4}$ (LSCO)\textsuperscript{\cite{zhou2004dichotomy}} and Tl$_{2}$Ba$_2$CuO$_{6+\delta}$ (Tl2201)\textsuperscript{\cite{plate2005fermi}}, the scattering rate is maximal along the nodal direction and decreases with the momentum moving from the nodal to the antinodal regions. The scattering rate we observed in our heavily overdoped Bi2212 sample 
 exhibits a different momentum dependence from those observed in LSCO\textsuperscript{\cite{zhou2004dichotomy}} and Tl2201\textsuperscript{\cite{plate2005fermi}}. We note that the scattering rate we observed in our overdoped Bi2212 sample is much smaller than those observed in LSCO\textsuperscript{\cite{zhou2004dichotomy}} and Tl2201\textsuperscript{\cite{plate2005fermi}}. This indicates that the momentum dependence of the scattering rate we observed is more intrinsic to the overdoped cuprates. 

In the overdoped region, the superfluid density is found to decrease with the increasing hole doping.\textsuperscript{\cite{bovzovic2016dependence}} This naturally raises the question of whether there exist normal electrons in the superconducting state in the overdoped samples. The normal electrons, if they exist, may be present in the form of phase separation either in the real space or reciprocal space as shown in the STM measurement of the overdoped Bi2201 samples.\textsuperscript{\cite{tromp2023puddle,ye2023emergent}} In principle, ARPES can distinguish normal electrons from superconducting electrons because they behave differently in the superconducting state. The normal electrons will have a band that crosses the Fermi level and produce spectral weight at the Fermi level. On the other hand, the superconducting electrons will form a gap near the Fermi level and cause spectral weight suppression at the Fermi level. Under the condition that the superconducting gap is larger than the single-particle scattering rate,
the spectral weight at the Fermi level can be fully suppressed to zero. In this case, if there are normal electrons coexisting with the superconducting electrons, they will produce spectral weight at the Fermi level. It can then be used to determine whether there are normal electrons and the fraction of normal electrons in the superconducting state. 

Figure 4(c) shows the momentum dependence of the ratio of the EDC intensity at the Fermi energy, $I_\mathrm{E_F}$ to that at the peak position, $I_\mathrm{Peak}$ in the symmetrized EDC, as illustrated in the inset of Fig. 4(c). The ratio is 1 along the nodal direction, decreases as the momentum moves away from the nodal direction and approaches nearly zero near the antinodal region. For superconducting electrons, the ratio depends on the relative magnitude of the superconducting gap ($\mathit{\Delta}$) and the single particle scattering rate ($\mathit{\Gamma}_1$). The ratio is 1 because the superconducting gap is zero along the nodal direction. Away from the nodal direction, the superconducting gap increases rapidly while the scattering rate changes slowly, giving rise to the decrease of the ratio $I_\mathrm{E_F}$/$I_\mathrm{Peak}$. Upon approaching the antinodal region, the superconducting gap becomes larger than the scattering rate and a hard gap is formed around the Fermi level where the spectral weight is fully suppressed to zero. These results indicate that in the momentum space, except for the nodal region, no other regions are observed to have a zero superconducting gap. This rules out the possibility of phase separation in the reciprocal space around the antinodal region in our heavily overdoped Bi2212 sample. 

Our present results can also rule out the possibility of phase separation in real space in our heavily overdoped Bi2212 sample. This is made possible because we can observe a hard gap opening near the antinodal region where the scattering rate is smaller than the superconducting gap. Three factors are essential to this observation: a clean sample, ultra-low temperature and ultra-high instrumental resolution. If the superconducting phase has a hard gap with the spectral weight at the Fermi level fully suppressed to zero, the presence of the normal phase can be detected because it will produce extra spectral weight at the Fermi level. As shown in Fig. 3(c) and Fig. 4(c), the spectral weight at the Fermi level near the antinodal region is nearly zero. This indicates that the normal phase is negligible in the superconducting state in our heavily overdoped Bi2212 sample.

Our present results indicate that overdoped Bi2212 behaves differently from the overdoped Bi2201. The normal fluid is clearly observed by STM in the superconducting state of the overdoped Bi2201.\textsuperscript{\cite{tromp2023puddle,ye2023emergent}} But in heavily overdoped Bi2212, the normal fluid is negligible in the superconducting state. In overdoped LSCO, it is found that the superfluid density decreases with increasing doping.\textsuperscript{\cite{bovzovic2016dependence}} In overdoped Bi2212, whether the superfluid density follows the same doping evolution remains to be investigated. Our observation of the absence of normal fluid in the superconducting state of the heavily overdoped Bi2212 can be understood if the superfluid density keeps increasing with increasing doping in overdoped Bi2212.\textsuperscript{\cite{hwang2007doping}} In that case, it provides important information to understand the origin of superconductivity in the overdoped region. 

~\\
\noindent{\bf\large{Summary}}
 
In summary, we have carried out high-resolution ultra-low temperature laser-based ARPES measurements on the heavily overdoped Bi2212 sample with a $T_\mathrm{c}$ of 48\,K.
We find that this heavily overdoped Bi2212 remains in the strong coupling regime with $2 \mathit{\Delta}_0 / k_{\mathrm{B}} T_{\mathrm{c}}=5.8$.
The single-particle scattering rate is very small along the nodal direction ($\sim$5\,meV) and increases as the momentum moves from the nodal to the antinodal regions.
A hard superconducting gap opening is observed near the antinodal region with the spectral weight at the Fermi level fully suppressed to zero. The normal fluid is found to be negligibly small in the superconducting state in this heavily overdoped Bi2212 sample. 
These results provide key information to understand the high $T_\mathrm{c}$ mechanism in the cuprate superconductors.

~\\

\vspace{3mm}

\noindent {\bf\large Acknowledgements}

This work is supported by the National Natural Science Foundation of China (Grant Nos. 12488201, 12374066, 12074411 and 12374154), the National Key Research and Development Program of China (Grant Nos. 2021YFA1401800, 2022YFA1604200, 2022YFA1403900 and 2023YFA1406000), the Strategic Priority Research Program (B) of the Chinese Academy of Sciences (Grant No. XDB25000000 and XDB33000000), the Innovation Program for 
Quantum Science and Technology (Grant No. 2021ZD0301800), the Youth Innovation Promotion Association of CAS (Grant No. Y2021006) and the Synergetic Extreme Condition User Facility (SECUF).

\vspace{3mm}

\noindent {\bf\large Author Contributions}

X.J.Z., X.T.L., L.Z. and C.H.Y. proposed and designed the research. 
C.H.Y. contributed to the sample growth and 
the magnetic measurements. 
C.H.Y., Y.Y.X., Y.W.C., J.G.Y., J.J.J., X.Z., W.K.L., H.T.Y., H.T.R., 
 S.J.Z., Z.M.W., N.Z., L.J.L., R.K.L., X.Y.W., F.F.Z., F.Y., Q.J.P., Z.Y.X., G.D.L., H.Q.M., L.Z., and X.J.Z. contributed to the development and maintenance of the ARPES systems and related 
software development. 
Q.H.W., J.H.L., X.Z., H.Q.M. and X.T.L. contributed to the development and maintenance of the STM. 
C.H.Y. and Q.H.W. carried out the ARPES and STM experiments with 
Y.Y.X., J.H.L., J.G.Y., J.J.J., X.Z., W.K.L., L.Z., X.T.L. and X.J.Z.. 
C.H.Y., Q.H.W., Y.W.C., J.H.L., L.Z., X.T.L. and X.J.Z. analysed the data. 
X.J.Z., L.Z., X.T.L. and C.H.Y. wrote the paper. All authors 
participated in discussion and commented on the paper.

\vspace{3mm}

\noindent{\bf\large Competing interests}\\
The authors declare no competing interests.

\newpage

\begin{figure*}[tbp]
\begin{center}
\includegraphics[width=1.0\columnwidth,angle=0]{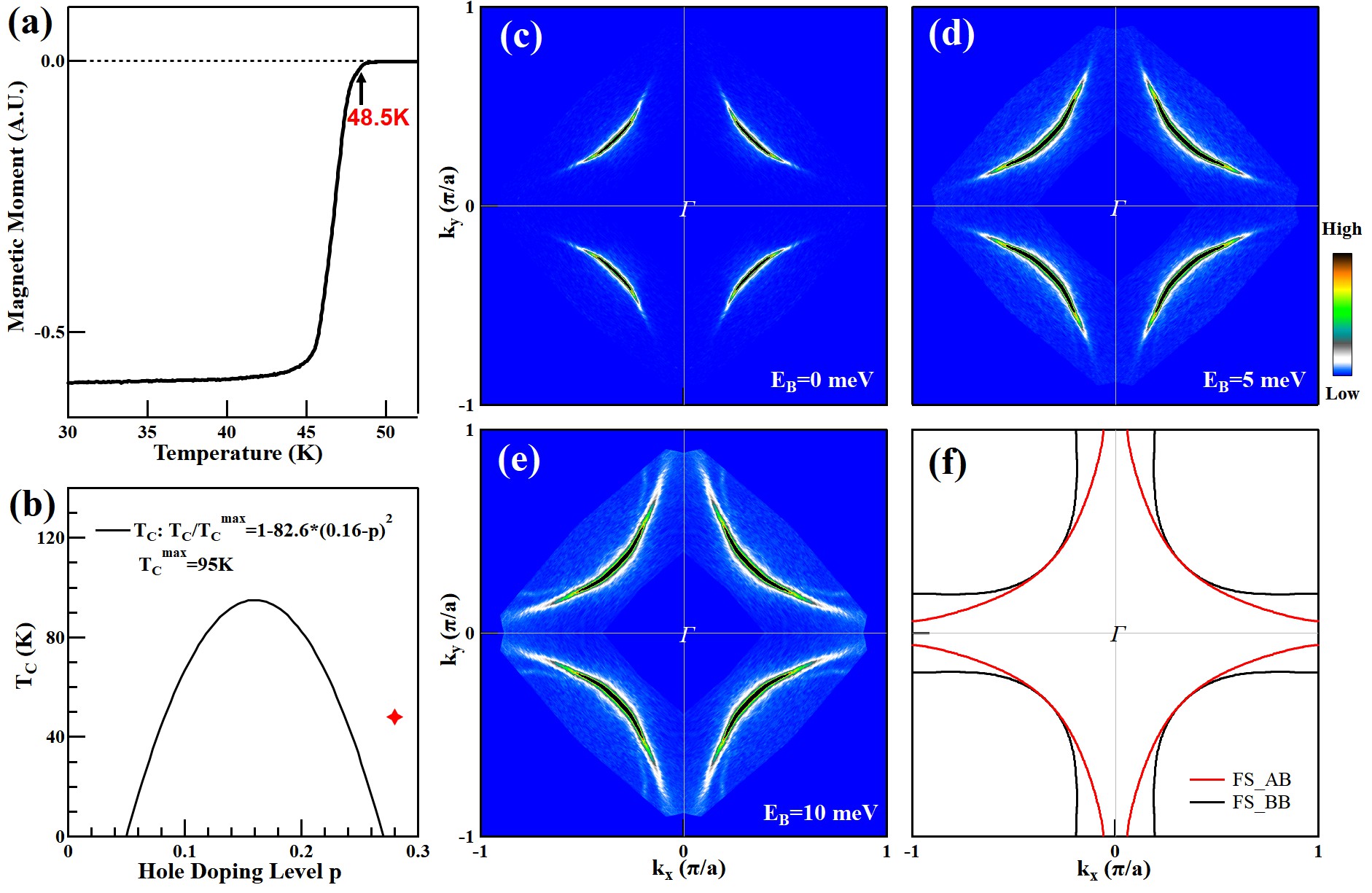}
\end{center}
\caption{\textbf{Fermi surface of the heavily overdoped Bi2212 with a $T_\mathrm{c}$$\sim$48.5\,K was measured at 1.2\,K.} \textbf{(a)} Magnetic susceptibility of the Bi2212 sample measured  under a magnetic field of 5\,Oe. The onset $T_\mathrm{c}$ is 48.5\,K with a transition width of $\sim$2\,K (10\%-90\%). \textbf{(b)} Schematic phase diagram of Bi2212 with a maximum at $T_\mathrm{c}$ at 95\,K.\textsuperscript{\cite{presland1991general}} The doping level of our measured sample is marked as an asterisk. \textbf{(c)-(e)} Fermi surface mapping (c) and constant energy contours at binding energies of 5\,meV (d) and 10\,meV (e). These images are obtained by integrating the spectral intensity within $\pm$1\,meV relative to the specified binding energy and are symmetrized by considering the fourfold symmetry of the main bands. \textbf{(f)} The measured Fermi surface of the sample. It consists of the antibonding ($\mathrm{FS \_ AB}$, red lines) and bonding ($\mathrm{FS \_ BB}$, black lines) Fermi surface sheets.
}
\end{figure*}

\begin{figure*}[tbp]
\begin{center}
\includegraphics[width=1.0\columnwidth,angle=0]{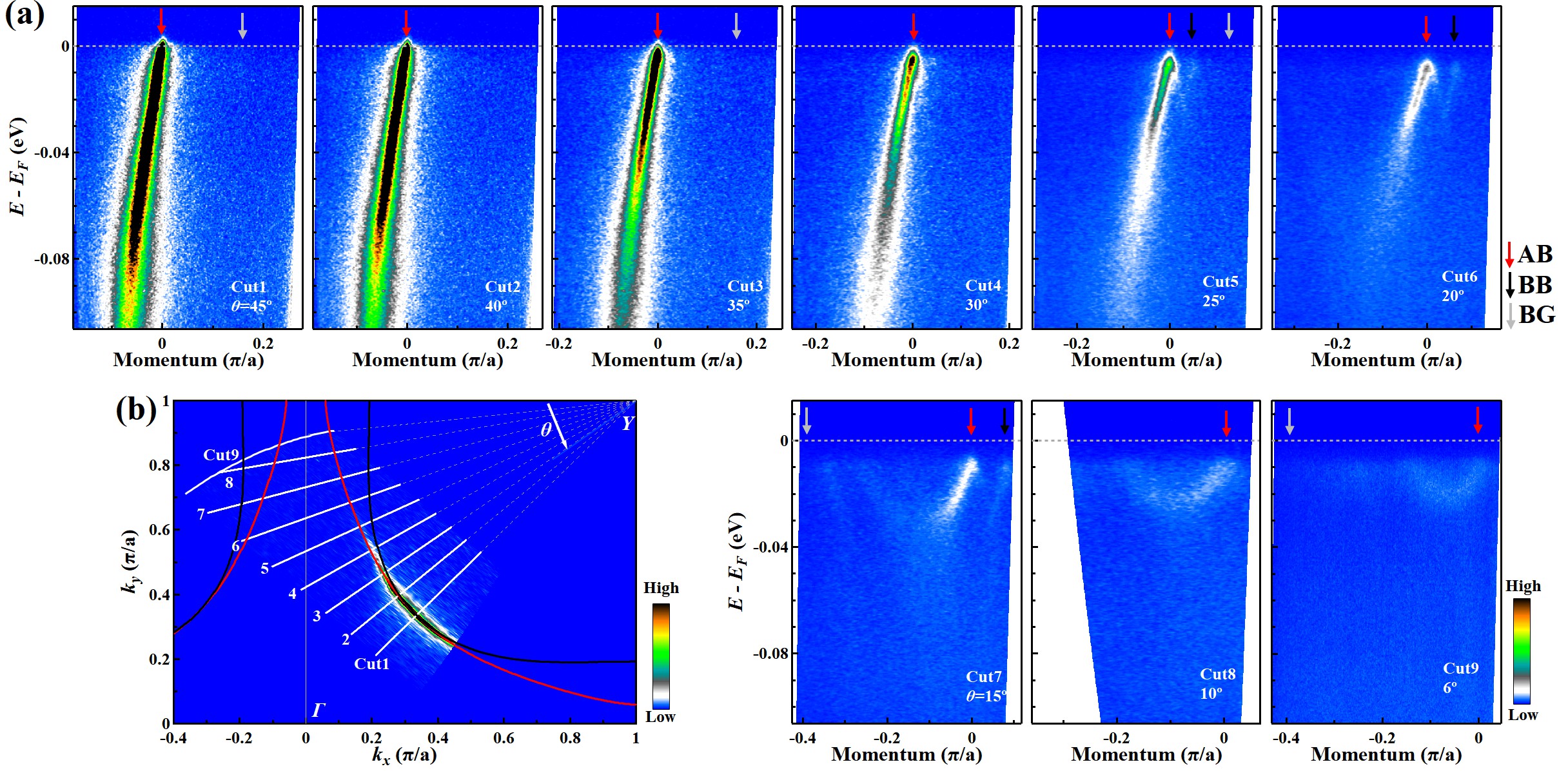}
\end{center}
\caption{\textbf{Momentum-dependent band structures of the Bi2212OD48K measured at 1.2\,K.} \textbf{(a)} Band structures measured along different momentum cuts from the nodal region to the antinodal region. The position of the momentum cuts is indicated in (b) by thick white lines. The location of the momentum cuts is also represented by the angle $\theta$ which is defined in (b). The red (black) arrows indicate the Fermi momentum position of the antibonding (bonding) bands while the grey arrows indicate the position without obvious bands. \textbf{(b)} The Fermi surface mapping of the Bi2212OD48K sample with the momentum cuts marked. 
}
\end{figure*}

\begin{figure*}[tbp]
\begin{center}
\includegraphics[width=1.0\columnwidth,angle=0]{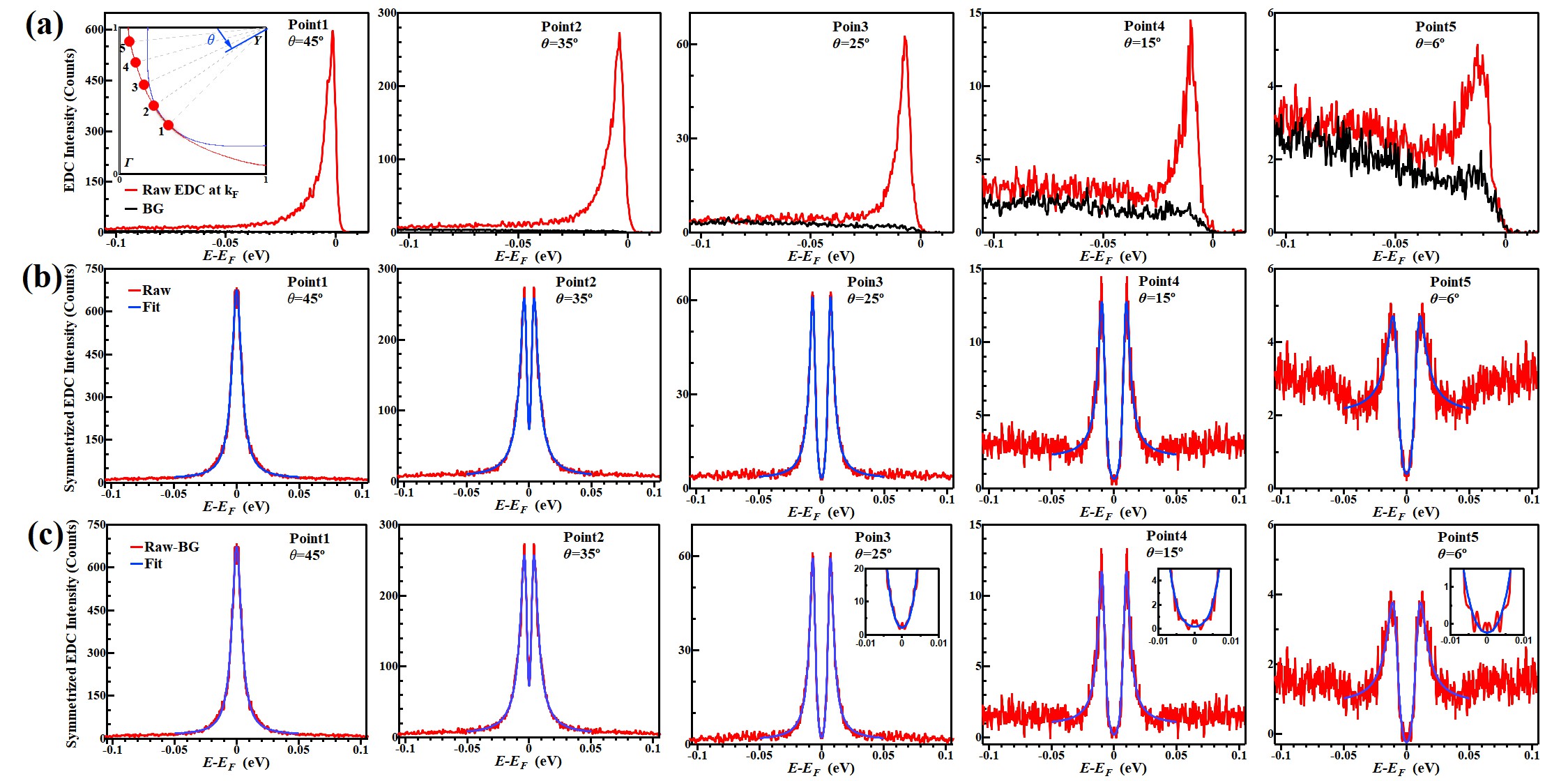}
\end{center}
\caption{\textbf{Representative photoemission spectra of Bi2212OD48K measured at 1.2\,K along the antibonding Fermi surface.} \textbf{(a)} EDCs measured at five different Fermi momenta (red lines). The position of the Fermi momenta are marked with red solid circles in the inset of the leftmost panel. The EDCs taken away from the Fermi surface are plotted as black lines which serve as background signal. Their momentum positions are marked by the grey arrows in Fig. 2(a). \textbf{(b)} The corresponding symmetrized EDCs at different Fermi momenta (red lines) obtained from (a). They are fitted by the phenomenological gap formula\textsuperscript{\cite{norman1998}} and the fitted curves are plotted by the blue lines. \textbf{(c)} Same as (b) but for the background-subtracted EDCs. The background-subtracted EDCs are obtained from (a) by subtracting the background EDCs from the EDCs at different Fermi momenta. The insets in the right three panels highlight the lower intensity region at the Fermi level. 
 }

\end{figure*}

\begin{figure*}[tbp]
	\begin{center}
		\includegraphics[width=1.0\columnwidth,angle=0]{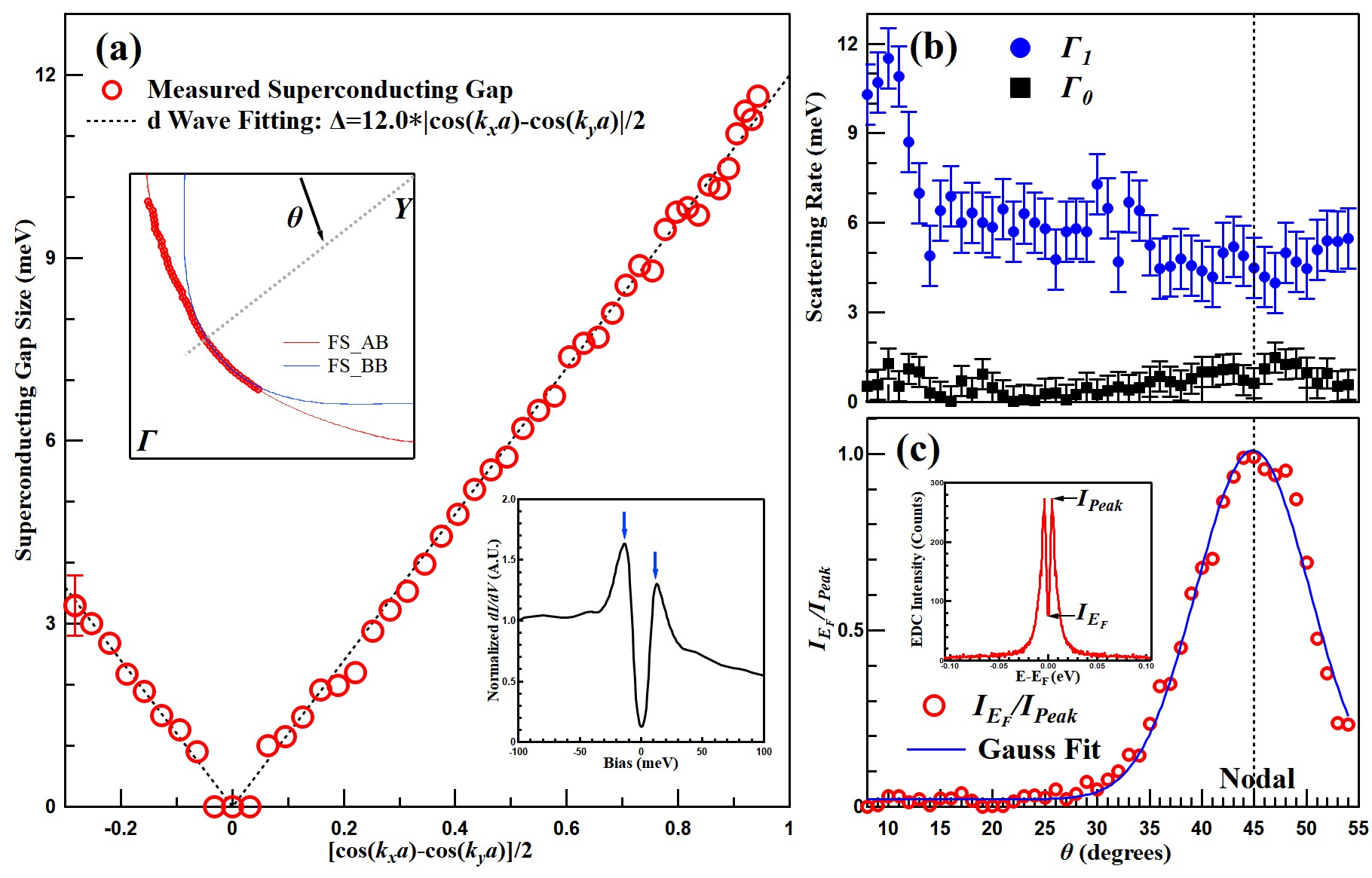}
	\end{center}
	\caption{\textbf{Momentum dependence of superconducting gap, scattering rate and normal fluid intensity in Bi2212OD48K} \textbf{(a)} The measured superconducting gap of the antibonding Fermi surface as a function of [cos($k_x$)-cos($k_y$)]/2. The dashed line represents a linear fit to the superconducting gap using a standard d-wave form $\mathit{\Delta} = \mathit{\Delta}_0 |\cos (k_x a)-\cos (k_y a)| / 2$. The top-left inset shows the schematic Fermi surface and definition of the Fermi surface angle $\theta$. The bottom-right inset shows the spatially averaged d$I$/d$V$ spectrum of the same Bi2212OD48K sample measured at 4.7\,K. The coherence peak lies at 12.5\,meV as marked by the blue arrows. \textbf{(b)} The momentum dependence of the scattering rates, $\mathit{\Gamma} _1$ (blue) and $\mathit{\Gamma} _0$ (black), along the antibonding Fermi surface. The scattering rates are obtained by fitting the symmetrized background-subtracted EDCs with the phenomenological gap formula\textsuperscript{\cite{norman1998}} as shown in Fig. 3(c). \textbf{(c)} The momentum dependence of normal fluid intensity. It is defined as the ratio of the EDC intensity at the Fermi level ($I_\mathrm{E_F}$) to the peak intensity ($I_\mathrm{Peak}$), as defined in the inset. The thick blue line represents the fitting result using a Gaussian function.
	} 	
\end{figure*}

\end{document}